\documentclass[onecolumn]{elsart}%
\usepackage{amsfonts}
\usepackage{amsmath}
\usepackage{amssymb}
\usepackage{graphicx}%
\providecommand{\U}[1]{\protect\rule{.1in}{.1in}}
\begin{document}
%
\begin{frontmatter}%
%

\title{Mixed Quantum States with Variable Planck Constant}%
%

\author{Maurice A. de Gosson}%
%

\address{University of Vienna, Faculty of Mathematics (NuHAG)
Email: maurice.de.gosson@univie.ac.at}%
%

\begin{abstract}%
Recent cosmological measurements tend to confirm that the fine structure
constant $\alpha$ is not immutable and has undergone a tiny variation since
the Big Bang. Choosing adequate units, this could also reflect a variation of
Planck's constant $h$. The aim of this Letter is to explore some consequences
of such a possible change of $h$ for the pure and mixed states of quantum
mechanics. Surprisingly enough it is found that not only is the purity of a
state extremely sensitive to such changes, but that quantum states can evolve
into classical states, and vice versa. A complete classification of such
transitions is however not possible for the moment being because of yet
unsolved mathematical difficulties related to the study of positivity
properties of trace class operators.%
\end{abstract}%
%

%

\end{frontmatter}%

\section{Introduction}

The variability of physical \textquotedblleft constants\textquotedblright\ is
a possibility that cannot be outruled and which has being an active area of
research for some time in cosmology and astrophysics. In fact, since Paul
Dirac \cite{Dirac1,Dirac} suggested in 1937 the \textquotedblleft Large
Numbers Hypothesis\textquotedblright\ that some constants of Nature could vary
in space and time, the topic has remained a subject of fascination which has
motivated numerous theoretical and experimental researches \cite{oklo1,Dyson1}.

The difficulty is not only of an experimental nature but also involves
delicate issues related to the choice of units. It is anyway problematic to
discuss the proposed rate of change (or lack thereof) of a single dimensional
physical constant in isolation \cite{Duff3}. The reason for this is that the
choice of a system of units may arbitrarily select any physical constant as
its basis, making the question of which constant is undergoing change an
artefact of the choice of units: these issues have to by studied in depth. In
the present Letter we begin by shortly discussing some recent advances on the
topic of varying natural constants; we thereafter focus on the quantum
mechanical consequences of possible changes in Planck's constant using the
Wigner formalism.

\section{On the Variability of Constants of Nature}

\subsection{The fine structure constant}

Some scientists have suggested that the fine structure constant $\alpha
=e^{2}/\hbar c$ might not be constant, but could vary over time and space. The
quest for testing this hypothesis is ongoing. The history actually started in
a quite romantic way, with the story of the Oklo natural nuclear reactor found
in a uranium mine in Central Africa in 1972 (see \cite{meshik,uzle} for
accounts of these findings). The measurements that were made in Oklo give
limits on variation of the fine-structure constant over the period since the
reactor was running that is \textit{ca}.1.8 billion years, which is much less
than the estimated age of Universe. In 1999, a team of astronomers headed by
J. Webb reported that measurements of light absorbed by very distant quasars
suggest that the value of the fine-structure constant was once slightly
different from what it is today. These experiments, made using Keck and VLT
telescopes in Hawaii, put an upper bound on the relative change per year, at
roughly $10^{-17}$ per year. In spite of many criticisms, Webb and his
collaborators seem to be progressing towards a confirmation of this variation
\cite{oklo2}. Anyway, as Feng and Yan \cite{FengYan} emphasize, space-time
variations of $\alpha$ in cosmology is a new phenomenon beyond the standard
model of physics which, if proved true, must mean that at least one of the
three fundamental constants $e,\hbar,c$ that constitute it must vary. This is
a delicate issue, related to choices of units, as will be discussed in a
moment. Also see Kraiselburd \textit{et al.} \cite{Krais} who analyze the
consistency of different astronomical data of the variation in the
fine-structure constant obtained with Keck and VLT telescopes.

\subsection{Planck's constant}

Planck's constant is a central number for modern physics, and it also appears
in the fine structure constant. To test whether Planck's constant is really
constant, Mohageg and Kentosh \cite{KM1,KM3} set out to measure possible
spatial discrepancies using the freely available data obtained from the same
GPS systems that car drivers use to find their way home. The story goes as
follows \cite{APS,bbc}: GPS relies on the most accurate timing devices we
currently possess: atomic clocks. These clocks count the passage of time
according to frequency of the radiation that atoms emit when their electrons
jump between different energy levels. Kentosh and Mohageg looked through a
year's worth of GPS data of seven highly stable GPS satellites and found that
the corrections depended in an unexpected way on a satellite's distance above
the Earth. This small discrepancy could be due to atmospheric effects or
random errors, but it could also arise from a position-dependent Planck's
constant. If $h$ changes from place to place, so do the frequencies, and thus
the \textquotedblleft ticking rate\textquotedblright, of atomic clocks. And
any dependence of $h$ on location would then translate as a tiny timing
discrepancy between different clocks. So, what did they discover? After
careful analysis of the data they obtained, Kentosh and Mohageg concluded that
$h$ is identical at different locations to an accuracy of seven parts in a
thousand. Their results, which have been largely commented (and criticized) in
the media, are however controversial; see for instance Berengut and Flambaum's
rebuttal \cite{befla}, and Kentosh and Mohageg's reply \cite{KM2}.

There are other recent work dealing with the possibility of a varying Planck's
constant. In \cite{SL} Seshavatharam and Lakshminarayana have discussed the
possibility of viewing Planck's constant as a cosmological variable; they
argue that using the cosmological rate of change in Planck's constant, the
future cosmic acceleration can be verified from the ground based laboratory
experiments. Mangano \textit{et al.} \cite{Mangano} consider the possibility
that the Planck constant is characterized by stochastic fluctuations. Their
study is motivated by Dirac's suggestion that fundamental constant are
dynamical variables and by conjectures on quantum structure of spacetime at
small distances; they assume that there is a time-dependence of $h$ consisting
in Gaussian random fluctuations around its constant mean value, with a typical
correlation time scale.

We notice that the quantity $h$ is also fundamental in the \textquotedblleft
positive\textquotedblright\ sense \cite{duffetal}: it is the quantum of the
angular momentum $J$ and a natural unit of the action $S$. When $J$ or $S$ are
close to $\hbar$, the whole realm of quantum mechanical phenomena appears
(this is very important because it makes clear the relationship between $h$
and minimal action and the notion of quantum blob we have developed elsewhere
\cite{gophysa,goletta,blobs}).

\subsection{Choice of units: a delicate problem}

M. Duff has remarked \cite{duffetal,Duff2,Duff3} that all the fundamental
physical dimensions could be expressed using only one unity: mass. Duff first
noticed the obvious, namely that length can be expressed as times using $c$,
the velocity of light, as a conversion factor. One can therefore take $c=1$,
and measure lengths in seconds. The second step is to use the relation
$E=h\nu$ which relates energy to a frequency, that is to the inverse of a
time. We can thus measure a time using the inverse of energy. But energy is
equivalent to mass as shown by Einstein, so that time can be measured by the
inverse of mass. Thus, setting $c=h=1$ we have reduced all the fundamental
dimensions to one: mass. A further step consists in choosing a reference mass
such that the gravitational constant is equal to one: $G=1$. Summarizing, we
have got a theoretical system of units in which $c=h=G=1$. There are other
ways to define irreducible unit systems. Already Stoney \cite{Stoney}, noting
that electric charge is quantized, derived units of length, time, and mass in
1881 by normalizing $G,c,$ and $e$ to unity; see the Wikipedia article
\cite{wiki}for a complete review of standard choices of units.

\section{The Dependency of Density Matrices on Planck's Constant}

\subsection{The $\eta$-Wigner distribution}

A mixed quantum state is the datum of a countable set of pairs $\mathcal{S}%
=\{(|\psi_{j}\rangle,\alpha_{j}):j\in\mathbb{N}\}$ where the $\psi_{j}$ are
normalized elements of some Hilbert space $\mathcal{H}$ and the $\alpha_{j}$
are positive real numbers playing the role of probabilities: $%
{\textstyle\sum_{j}}
\alpha_{j}=1$. The datum of $\mathcal{S}$ is equivalent to that of the density
matrix%
\begin{equation}
\widehat{\rho}=\sum_{j}\alpha_{j}|\psi_{j}\rangle\langle\psi_{j}|;
\label{rho1}%
\end{equation}
the $\widehat{\rho}_{j}=|\psi_{j}\rangle\langle\psi_{j}|$ are the orthogonal
projectors on the ray generated by $\psi_{j}$; they are identified with the
pure states $|\psi_{j}\rangle$. Let us now be more specific, and choose once
for all $\mathcal{H}=L^{2}(\mathbb{R}^{n})$ (the square integrable functions
on the configuration space $\mathbb{R}^{n}$). To the density matrix
$\widehat{\rho}$ let us associate the $\eta$-Wigner distribution
\begin{equation}
P_{\eta}(x,p)=\left(  \tfrac{1}{2\pi\eta}\right)  ^{n}\int\langle x+\tfrac
{1}{2}y|\widehat{\rho}|x-\tfrac{1}{2}y\rangle e^{-\frac{i}{\eta}py}d^{n}y
\label{peta}%
\end{equation}
where $\eta$ is a non-zero real parameter. For the choice $\eta=\hbar$ the
function $P_{\hbar}=P$ is just the usual Wigner distribution
\cite{Wigner,Hillery} commonly used in quantum mechanics. Using the definition
(\ref{rho1}) of the density matrix, we can rewrite formula (\ref{peta}) in the
more explicit form%
\begin{equation}
P_{\eta}(x,p)=\sum_{j}\alpha_{j}W_{\eta}\psi_{j}(x,p) \label{petaeta}%
\end{equation}
where
\begin{equation}
W_{\eta}\psi_{j}(x,p)=\left(  \tfrac{1}{2\pi\eta}\right)  ^{n}\int
e^{-\frac{i}{\eta}py}\psi_{j}(x+\tfrac{1}{2}y)\psi_{j}^{\ast}(x-\tfrac{1}%
{2}y)d^{n}y \label{wigner1}%
\end{equation}
is the $\eta$-Wigner transform of $\psi_{j}$; when $\eta=\hbar$ we recover the
usual Wigner transform $W_{\hbar}\psi_{j}=W\psi_{j}$.

Exactly as the operator $\widehat{\rho}$ is the Weyl transform of $P$, it is
also the $\eta$-Weyl transform of $P_{\eta}$ in the sense that $\widehat{\rho
}$ has the harmonic decomposition%
\begin{equation}
\widehat{\rho}=\iint F_{\eta}\rho(x,p)e^{-\frac{i}{\eta}(x\widehat
{x}+p\widehat{p})}d^{n}xd^{n}p \label{weyl1}%
\end{equation}
where $F_{\eta}\rho$ is the $\eta$-Fourier transform of $\rho$:%
\begin{equation}
F_{\eta}\rho(x,p)=\left(  \tfrac{1}{2\pi\eta}\right)  ^{n}\iint e^{-\frac
{i}{\eta}(xx^{\prime}+pp^{\prime})}\rho(x^{\prime},p^{\prime})d^{n}x^{\prime
}d^{n}p^{\prime}. \label{ft}%
\end{equation}
Formula (\ref{weyl1}) shows that $(2\pi\eta)^{n}\rho$ is the Weyl symbol of
the operator $\widehat{\rho}$ \cite{Birk,Birkbis,Littlejohn}, that is, we have%
\begin{equation}
\widehat{\rho}\psi(x)=\iint e^{\frac{i}{\eta}p(x-y)}\rho(\tfrac{1}%
{2}(x+y),p)\psi(y)d^{n}yd^{n}p \label{Weyl2}%
\end{equation}
for every wavefunction $\psi$.

We are allowing the parameter $\eta$ to take negative values. The change of a
positive $\eta$ to the negative value $-\eta$ has a simple physical
interpretation: it corresponds to a reversal of the arrow of time. We have the
following explicit formula relating $W_{\eta}\psi$ and $W_{-\eta}\psi$:%
\begin{equation}
W_{\eta}\psi=(-1)^{n}W_{-\eta}(\psi^{\ast});\label{wmineta}%
\end{equation}
this is readily proved by making the substitution $y\longmapsto-y$ in the
integral in formula (\ref{wigner1}). An easy (but important) consequence of
this equality is that the marginal properties
\begin{equation}
\int W_{\eta}\psi(x,p)d^{n}x=|F_{\eta}\psi(p)|^{2}\text{ \ , \ }\int W_{\eta
}\psi(x,p)d^{n}p=|\psi(x)|^{2}\label{marginal}%
\end{equation}
hold for all the $\psi\in L^{2}(\mathbb{R}^{n})$ such that $\psi\in
L^{1}(\mathbb{R}^{n})$ and $F_{\eta}\psi\in L^{1}(\mathbb{R}^{n})$; here%
\[
F_{\eta}\psi(p)=\left(  \tfrac{1}{2\pi|\eta|}\right)  ^{n/2}\int e^{-\frac
{i}{\eta}px}\psi(x)d^{n}x.
\]
Another consequence is that the Moyal identity
\begin{equation}
\int W_{\eta}\psi(z)W_{\eta}\phi(z)d^{n}z=\left(  \tfrac{1}{2\pi|\eta
|}\right)  ^{n}|\langle\psi|\phi\rangle|^{2}\label{Moyal}%
\end{equation}
holds for all square integrable functions $\psi$ and $\phi$ and all $\eta
\neq0$ (we are writing for short $z=(x,p)$). In fact formula (\ref{Moyal}) is
well known \cite{Wigner,Hillery} when $\eta>0$ (it suffices to replace $\hbar$
with $\eta>0$ in the proof); when $\eta<0$ formula (\ref{wmineta}) shows that%
\[
\int W_{\eta}\psi(z)W_{\eta}\phi(z)d^{n}z=\int W_{-\eta}\psi^{\ast}%
(z)W_{-\eta}\phi^{\ast}(z)d^{n}z
\]
and since $-\eta>0$ this leads us back to the former case:%
\[
\int W_{-\eta}\psi^{\ast}(z)W_{-\eta}\phi^{\ast}(z)d^{n}z=\left(  \tfrac
{1}{2\pi|\eta|}\right)  ^{n}|\langle\psi|\phi\rangle|^{2}%
\]
since $\langle\psi^{\ast}|\phi^{\ast}\rangle=\langle\psi|\phi\rangle^{\ast}$.
Notice that the Moyal identity implies that the functions $\psi$ and $\phi$
are orthogonal if and only if their $\eta$-Wigner transforms are.

\subsection{Transitions of quantum states}

Consider the following situation: we have an unknown quantum system, on which
we perform a quorum of measurements (for instance, by a homodyne quantum
tomography \cite{Vogel}) in order to determine its (quasi)probability
distribution $\rho(z)=\rho(x,p)$. The latter allows us then to infer the
density matrix $\widehat{\rho}$ using the Weyl correspondence (\ref{Weyl2}).
It should however be clear that the result will depend on the value of
Planck's constant. Suppose that $\rho(z)$ is the Wigner distribution in the
usual sense (\textit{i.e.} with $\eta=\hbar$) of a density matrix. Then there
exists normalized square integrable functions $\psi_{1},\psi_{2},...$ and
positive constants $\alpha_{1},\alpha_{2},...$ summing up to one and such that%
\begin{equation}
\rho(z)=\sum_{j}\alpha_{j}W\psi_{j} \label{rhoh}%
\end{equation}
and $\rho(z)$ is thus the Wigner distribution of a density matrix
$\widehat{\rho}$. Can $\rho(z)$ be the $\eta$-Wigner distribution of another
density operator $\widehat{\rho}_{\eta}$ that is, can we have
\begin{equation}
\rho(z)=\sum_{j}\beta_{j}W_{\eta}\phi_{j} \label{rhoeta}%
\end{equation}
where the $\phi_{j}$ are normalized and\ $\beta_{j}\geq0$, $\sum_{j}\beta
_{j}=1$ ? We are going to see that this indeed possible, but only if some
severe conditions are imposed to the probabilities $\beta_{j}$. We begin by
making a remark that will considerably simplify our argument. If
$\widehat{\rho}$ is a density matrix, it is a trace class operator and is
hence compact. But it then follows from the spectral decomposition theorem
that $\rho(z)$ can be written in the form $\rho(z)=\sum_{j}\alpha_{j}^{\prime
}W\psi_{j}^{\prime}$ with $\alpha_{j}^{\prime}\geq0$, $\sum_{j}\alpha
_{j}^{\prime}=1$ and the vectors $\psi_{j}^{\prime}$ forming an orthonormal
system (the $\alpha_{j}^{\prime}$ are the eigenvalues of $\widehat{\rho}$ and
the $\psi_{j}^{\prime}$ are the corresponding normalized eigenfunctions);
there is thus no restriction to assume that the vectors $\psi_{j}$ in
(\ref{rhoh}) are orthonormal. The same argument applies to $\widehat{\rho
}_{\eta}$ so we can also assume that the vectors $\phi_{j}$ are orthonormal.
So let us assume that%
\begin{equation}
\sum_{j}\alpha_{j}W\psi_{j}=\sum_{j}\beta_{j}W_{\eta}\phi_{j}; \label{abcp}%
\end{equation}
squaring both sides of this equality we get%
\begin{equation}
\sum_{j,k}\alpha_{j}\alpha_{k}W\psi_{j}W\psi_{k}=\sum_{j,k}\beta_{j}\beta
_{k}W_{\eta}\phi_{j}W_{\eta}\phi_{k}. \label{square}%
\end{equation}
Now, by the Moyal formula (\ref{Moyal}) we have%
\begin{gather*}
\int W\psi_{j}(z)W\psi_{k}(z)d^{n}z=\left(  \tfrac{1}{2\pi\hbar}\right)
^{n}\delta_{jk}\\
\int W_{\eta}\phi_{j}(z)W_{\eta}\phi_{k}(z)d^{n}z=\left(  \tfrac{1}{2\pi
|\eta|}\right)  ^{n}\delta_{jk}%
\end{gather*}
hence, integrating both sides of the equality (\ref{square}), we are led to
the condition
\begin{equation}
\left(  \tfrac{1}{2\pi\hbar}\right)  ^{n}\sum_{j}\alpha_{j}^{2}=\left(
\tfrac{1}{2\pi|\eta|}\right)  ^{n}\sum_{j}\beta_{j}^{2}. \label{traceq}%
\end{equation}
This equality can be interpreted in terms of the purity $\operatorname*{Tr}%
(\widehat{\rho}^{2})$ and $\operatorname*{Tr}(\widehat{\rho_{\eta}}^{2})$ of
the states $\widehat{\rho}=\sum_{j}\alpha_{j}$ and $\widehat{\rho}_{\eta}$. In
fact, since the vectors $\psi_{j}$ are orthonormal it follows from formula
(\ref{rho1}) that%
\begin{align*}
\widehat{\rho}^{2}  &  =\left(
{\textstyle\sum\nolimits_{j}}
\alpha_{j}|\psi_{j}\rangle\langle\psi_{j}|\right)  ^{2}\\
&  =%
{\textstyle\sum\nolimits_{j,k}}
\alpha_{j}\alpha_{k}|\psi_{j}\rangle\langle\psi_{j}|\psi_{k}\rangle\langle
\psi_{k}|\\
&  =%
{\textstyle\sum\nolimits_{j}}
\alpha_{j}^{2}|\psi_{j}\rangle\langle\psi_{j}|
\end{align*}
hence $\operatorname*{Tr}(\widehat{\rho}^{2})=%
{\textstyle\sum\nolimits_{j}}
\alpha_{j}^{2}$ and, similarly, $\operatorname*{Tr}(\widehat{\rho_{\eta}}%
^{2})=%
{\textstyle\sum\nolimits_{j}}
\beta_{j}^{2}$. The condition (\ref{traceq}) can therefore be rewritten in the
simple form
\begin{equation}
|\eta|^{n}\operatorname*{Tr}(\widehat{\rho}^{2})=\hbar^{n}\operatorname*{Tr}%
(\widehat{\rho_{\eta}}^{2}). \label{traceqbis}%
\end{equation}
This interesting formula shows that the purity of a mixed quantum state
crucially depends on the value of Planck's constant. For instance if
$\operatorname*{Tr}(\widehat{\rho})=\operatorname*{Tr}(\widehat{\rho_{\eta}%
}^{2})=1$ then we must have $|\eta|=\hbar$: no pure state remains pure if we
change Planck's constant (except for a change of sign in $h$ corresponding to
time reversal). More generally, if $\widehat{\rho}$ is a pure state
$|\psi\rangle\langle\psi|$ then formula (\ref{traceqbis}) becomes
$\operatorname*{Tr}(\widehat{\rho_{\eta}}^{2})=(|\eta|/\hbar)^{n}$ hence
$\widehat{\rho_{\eta}}$ can be a mixed quantum state only if $|\eta|<\hbar$
and any decrease of $|\eta|$ leads to a loss of purity.

\section{Gaussian Mixed States}

Let us study in some detail the possible transitions of Gaussian states; in
addition to their intrinsic interest and importance, Gaussian states are the
only one whose dependence on Planck's constant is fully understood for the
moment being (see however Dias and Prata's \cite{dipra} analysis of
non-Gaussian pure states).

\subsection{A very simple example}

Consider the centered normal probability distribution on $\mathbb{R}^{2}$
defined by
\begin{equation}
\rho_{X,P}(x,p)=\frac{1}{2\pi\sigma_{X}\sigma_{P}}\exp\left[  -\frac{1}%
{2}\left(  \frac{x^{2}}{\sigma_{X}^{2}}+\frac{p^{2}}{\sigma_{P}^{2}}\right)
\right]  \label{Gauss1}%
\end{equation}
where $\sigma_{X},\sigma_{P}>0$; the associated variances are $\sigma_{X}^{2}$
and $\sigma_{P}^{2}$ and the covariance is $\sigma_{XP}=0$. Suppose now that
the value of Planck's constant is \textit{hic et nunc} $h$. It is known that
$\rho_{X,P}$ is the Wigner distribution of a density operator $\widehat{\rho
}_{X,P}$ if and only if it satisfies the Heisenberg inequality $\sigma
_{X}\sigma_{P}\geq\frac{1}{2}\hbar$ where $\hbar=h/2\pi$. If we have
$\sigma_{X}\sigma_{P}=\frac{1}{2}\hbar$ (which we assume from now on) then
$\rho_{X,P}$ is the Wigner distribution of the coherent state
\begin{equation}
\psi_{X}(x)=(2\pi\sigma_{X}^{2})^{-1/4}e^{-x^{2}/2\sigma_{X}^{2}} \label{psix}%
\end{equation}
and $\widehat{\rho}_{X,P}$ is then just the pure-state density matrix
$|\psi_{X}\rangle\langle\psi_{X}|$. Notice that $\hbar$ does not appear
explicitly in the function (\ref{psix}). Suppose now that we move the
distribution $\rho_{X,P}$ in space-time, to a location where $\hbar$ has a new
value $\eta>0$. Then $\sigma_{X}$ and $\sigma_{P}$ must satisfy the new
Heisenberg inequality $\sigma_{X}\sigma_{P}\geq\frac{1}{2}\eta$ to qualify
$\rho_{X,P}$ as a Wigner distribution, which implies that we must have
$\eta\leq\hbar$ since we have fixed $\sigma_{X}\sigma_{P}$ equal to $\frac
{1}{2}\hbar$. Physically this means that if we decrease the value of Planck's
constant then $\widehat{\rho}_{X,P}$ is the density operator of a (mixed)
quantum state, but if we increase its value so that $\eta>\hbar$ then the
Gaussian $\rho_{X,P}$ can only be viewed as the probability density of a
classical state -- it no longer represents a quantum state. In this case we
are witnessing a transition from the quantum world to the classical world.

\subsection{General multi-mode Gaussians}

Let us now consider Gaussians of the type
\begin{equation}
\rho_{\Sigma}(z)=(2\pi)^{-n}\sqrt{\det\Sigma^{-1}}e^{-\frac{1}{2}\Sigma
^{-1}z^{2}}%
\end{equation}
where $\Sigma$ is a positive definite symmetric (real) $2n\times2n$ matrix
(the \textquotedblleft covariance matrix\textquotedblright). We have $\rho
\geq0$ and
\[
\int\rho_{\Sigma}(z)d^{2n}z=1
\]
hence the function $\rho$ can always be viewed as a classical probability
distribution. It is the $\eta$-Wigner distribution of a density matrix if and
only if $\Sigma$ satisfies the positivity condition
\begin{equation}
\Sigma+\frac{i\eta}{2}J\geq0 \label{sigma1}%
\end{equation}
where $J=%
\begin{pmatrix}
0 & I\\
-I & 0
\end{pmatrix}
$ is the standard $2n\times2n$ symplectic matrix (this condition means that
all the eigenvalues of $\Sigma+(i\eta/2)J$ are $\geq0$). This can be proven in
several ways \cite{Dutta,Narcow3,goluPR}, but none of the available proofs is
really elementary. Condition (\ref{sigma1}) can be restated in several
different ways. The most convenient is to use the symplectic eigenvalues of
the covariance matrix. Observing that the product $J\Sigma$ has the same
eigenvalues as the antisymmetric matrix $\Sigma^{1/2}J\Sigma^{1/2}$ (because
they are conjugate) its eigenvalues are pure imaginary numbers $\pm
i\lambda_{1}^{\sigma},\pm i\lambda_{2}^{\sigma},...,\pm i\lambda_{n}^{\sigma}$
where $\lambda_{j}^{\sigma}>0$ for $j=1,2,...,n$. The set $\{\lambda
_{1}^{\sigma},\lambda_{2}^{\sigma},...,\lambda_{n}^{\sigma}\}$ is called the
\textit{symplectic spectrum} of $\Sigma$. Now, there exists a symplectic
matrix $S$ (\textit{i.e.} a matrix such that $S^{T}JS=J$) diagonalizing
$\Sigma$ as follows:
\begin{equation}
\Sigma=S^{T}DS\text{ \ , \ }D=%
\begin{pmatrix}
\Lambda & 0\\
0 & \Lambda
\end{pmatrix}
\label{Williamson}%
\end{equation}
where $\Lambda$ is the diagonal matrix with non-zero entries the positive
numbers $\lambda_{1}^{\sigma},\lambda_{2}^{\sigma},...,\lambda_{n}^{\sigma}$
(this is called a symplectic, or Williamson, diagonalization of $\Sigma$
\cite{Dutta,Birk,goluPR}). Since $S^{T}JS=J$ we have%
\[
\Sigma+\frac{i\eta}{2}J=S^{T}DS+\frac{i\eta}{2}J=S^{T}(D+\frac{i\eta}{2}J)S
\]
hence the condition $\Sigma+\frac{i\eta}{2}J\geq0$ is equivalent to
$D+\frac{i\eta}{2}J\geq0$. Now, the characteristic polynomial of the matrix
$D+\frac{i\eta}{2}J$ is the product $P_{1}(\lambda)\cdot\cdot\cdot
P_{n}(\lambda)$ where the $P_{j}$ are the second degree polynomials
$P_{j}(\lambda)=(\lambda_{j}^{\sigma}-\lambda)^{2}-\tfrac{\eta^{2}}{4}$ hence
the eigenvalues $\lambda$ of $D+\frac{i\eta}{2}J$ are the numbers
$\lambda=\lambda_{j}^{\sigma}\pm\frac{1}{2}\eta$; the condition $D+\frac
{i\eta}{2}J\geq0$ implies that all these eigenvalues $\lambda_{j}$ must be
$\geq0$ and hence $\lambda_{j}^{\sigma}\geq\sup\{\pm\frac{1}{2}\eta\}=\frac
{1}{2}|\eta|$ for all $j$. We have thus proven the equivalence
\begin{equation}
\Sigma+\frac{i\eta}{2}J\geq0\Longleftrightarrow|\eta|\leq2\lambda_{\min
}^{\sigma} \label{equiv}%
\end{equation}
where $\lambda_{\min}^{\sigma}$ is the smallest symplectic eigenvalue of the
covariance matrix $\Sigma$. The purity of the corresponding $\eta$-density
matrix is (\cite{Birk}, p. 302)%
\begin{equation}
\operatorname*{Tr}(\widehat{\rho_{\Sigma}}_{\eta}^{2})=\left(  \frac{\eta}%
{2}\right)  ^{n}\det(\Sigma^{-1/2}) \label{puresigma}%
\end{equation}
hence $\widehat{\rho_{\Sigma}}_{\eta}$ is a pure state if and only if
$\det(\Sigma)=(\eta/2)^{n}$. Since $\det(\Sigma)=\det(J\Sigma)=(\lambda
_{1}^{\sigma})^{2}\cdot\cdot\cdot(\lambda_{n}^{\sigma})^{2}$ this requires
that $\lambda_{j}^{\sigma}=1$ for all $j=1,2,...,n$ in view (\ref{equiv}). In
this case the matrix $D$ in (\ref{Williamson}) is the identity and
$\Sigma=S^{T}S$; the corresponding state is then a squeezed coherent state
\cite{Birk,Birkbis,goluPR,Littlejohn}; namely the image of the fiducial
coherent state $\phi_{0}(x)=(\pi\hbar)^{-n}e^{-|x|^{2}/2\hbar}$ by any of the
two metaplectic operators $\pm\widehat{S}$ defined by the symplectic matrix
$S$. To summarize, we have the following situation (we assume here for
simplicity that $\eta>0$): suppose that (\ref{sigma1}) holds for $\eta=\hbar$.
Then the system is a mixed quantum state for all $\eta\leq\hbar$; when
$\hbar\leq\eta\leq2\lambda_{\min}^{\sigma}$ it is still a mixed state unless
$\eta=\lambda_{1}^{\sigma}=\cdot\cdot\cdot=\lambda_{n}^{\sigma}$ in which case
it becomes a coherent state; when $\eta>2\lambda_{\min}^{\sigma}$ we are in
the presence of a classical Gaussian state.

\section{Discussion}

We have given necessary conditions for a quantum state to remain a quantum
state if Planck's constant undergoes a variation. To find sufficient
conditions is a very difficult mathematical problem related to the study of
positivity of trace class operators .Theoretical conditions allowing to test
the positivity of a given trace class operators were actually developed by
Kastler \cite{Kastler}, Loupias and Miracle-Sole
\cite{LouMiracle1,LouMiracle2} in the late 1960s using the theory of $C^{\ast
}$-algebras, and further studied by Narcowich and O'Connell
\cite{Narcow2,Narcow3,Narconnell,Narconnell88} and Werner \cite{Werner}. These
conditions (the \textquotedblleft KLM conditions\textquotedblright)\ are
however of limited practical use and no major advances have been made since
then (see however the paper of Dias and Prata \cite{dipra} which deals with
pure states). Very little is actually known about the consequences of a
varying Planck constant outside the Gaussian case we discussed above; it can
be shown that while the condition
\begin{equation}
\Sigma+\frac{i\hbar}{2}J\geq0\label{dutta}%
\end{equation}
is necessary for a trace class operator $\widehat{\rho}$ to be positive (and
hence to be a quantum state), it is not sufficient. In fact, this condition
alone (which is equivalent to the Robertson--Schr\"{o}dinger inequalities)
does not ensure positivity \cite{golubis,goluPR}, except in the Gaussian case
studied above (there are counterexamples where (\ref{dutta}) is satisfied
while the corresponding operator is non-positive
\cite{Narcow2,Narcow3,Manko,golubis}). We are discussing in a new work
\cite{cogoni17} alternative methods to study these issues using the theory of
Weyl--Heisenberg frames. These methods might provide a better insight in these
difficult questions.

\begin{ack}
The author has been financed by the Grant P27773 of the Austrian Science
Foundation FWF.
\end{ack}

\end{document}